\documentclass{article}

\usepackage{arxiv}

\usepackage{amsmath}
\usepackage{amssymb}
\usepackage{svg}
\usepackage{bbm}
\usepackage{makecell}

\usepackage[utf8]{inputenc} % allow utf-8 input
\usepackage[T1]{fontenc}    % use 8-bit T1 fonts
\usepackage{hyperref}       % hyperlinks
  \usepackage{cleveref}
\usepackage{url}            % simple URL typesetting
\usepackage{booktabs}       % professional-quality tables
\usepackage{amsfonts}       % blackboard math symbols
\usepackage{nicefrac}       % compact symbols for 1/2, etc.
\usepackage{microtype}      % microtypography
\usepackage{graphicx}
\graphicspath{ {./images/} }

\newcommand{\pyUAT}{\texttt{PyUAT}}
\newcommand{\Cglut}{\textit{C.~glutamicum}}

\newcommand{\UAT}{UAT}
\newcommand{\twalks}{\texttt{tensor\_walks}}

\author{
 Johannes Seiffarth \\
  Institute of Bio- and Geosciences, IBG-1: Biotechnology, \\
  Forschungszentrum Jülich, Jülich, Germany \\
  and \\
  Computational Systems Biotechnology (AVT.CSB), \\
  RWTH Aachen University, Aachen, Germany \\
   \And
 Katharina Nöh \\
  Institute of Bio- and Geosciences, IBG-1: Biotechnology, \\
  Forschungszentrum Jülich, Jülich, Germany \\
  \texttt{k.noeh@fz-juelich.de}
}

\begin{document}

\title{PyUAT: Open-source Python framework for efficient and scalable cell tracking with uncertainty awareness}

\date{\vspace{-5ex}}

\maketitle

\begin{abstract}
Tracking individual cells in live-cell imaging provides fundamental insights, inevitable for studying causes and consequences of phenotypic heterogeneity, responses to changing environmental conditions or stressors. Microbial cell tracking, characterized by stochastic cell movements and frequent cell divisions, remains a challenging task when imaging frame rates must be limited to avoid counterfactual results. A promising way to overcome this limitation is uncertainty-aware tracking (UAT), which uses statistical models, calibrated to empirically observed cell behavior, to predict likely cell associations. We present \pyUAT{}, an efficient and modular Python implementation of UAT for tracking microbial cells in time-lapse imaging. We demonstrate its performance on a large 2D+t data set and investigate the influence of modular biological models and imaging intervals on the tracking performance.
The open-source \pyUAT{} software is available at \url{https://github.com/JuBiotech/PyUAT}, including example notebooks for immediate use in Google Colab.
\end{abstract}

\date{\vspace{-5ex}}

% keywords can be removed

%%%%%%%%%%%%%%%%%%%%%%%%%%%%%%%%%%%%%%%%%%%%%%
%%%%%%%%%%%%%%%%%%%%%%%%%%%%%%%%%%%%%%%%%%%%%%
\section{Introduction}
\label{sec:intro}
%%%%%%%%%%%%%%%%%%%%%%%%%%%%%%%%%%%%%%%%%%%%%%
%%%%%%%%%%%%%%%%%%%%%%%%%%%%%%%%%%%%%%%%%%%%%%
Microfluidic live-cell imaging (MLCI) is an emerging high-throughput technology for monitoring the spatio-temporal development of microbial cells under precisely controllable conditions, with hundreds of replicates per experiment~\cite{Grunberger2012, Ugolini2024}. Because of its ability to record the development of individual cells within 2D monolayer cavities, MLCI is ideally suited for studying the causes and consequences of phenotypic heterogeneity that occurs within isogenic microbial populations and consortia. This capability has been proven to be highly informative, as evidenced by diverse applications in biomedical, biotechnological, and ecological fields~\cite{jeckel_advances_2021, Balaban2004, Micali2023}. For example, MLCI has provided unique quantitative insights into the phenotypic heterogeneity of microbial organisms in constant and fluctuating environments~\cite{mustafi_application_2014, kasahara_enabling_2023, Blobaum2023}, responses upon exposure to stress factors~\cite{helfrich_live_2015}, or the impact of biological noise~\cite{Delvigne2017}. 

To gain insight into the development of colonies, accurate cell pedigrees spanning several generations need to be extracted from the time-lapse images. This information is captured in cell lineage trees (CLT), which are bifurcated trees with the cell instances serving as nodes and the edges represent the frame-to-frame associations of the cells, with a branch indicating a cell division (\Cref{fig:workflow}).
The components of generating CLTs are, thus, the segmentation of individual cells in each image and the tracking of these cells throughout the time-lapse. Today, high-quality deep-learning (DL) segmentation models are available for the microbial domain, providing accurate results across organisms and imaging modalities~\cite{cutler_omnipose_2022, Seiffarth2024,Ma2024}.
Cell tracking, however, is generally considered to be a more complicated task, with DL-based solutions only recently proposed for 1D micro-channels (so-called mother machines) and 2D micro-chambers~\cite{OConnor2024,gallusser_trackastra_2024}. Still, in the microbial domain ground truth tracking data is rare. Therefore, classical (non-DL) linear assignment problem (LAP)-based trackers prevail, which predict the edges in a two-step procedure: first, the costs for potential edge candidates are determined from cell features (e.g. distance or mask overlap), and then edges with minimal costs are selected to link cells between frames. This process is repeated for each pair of subsequent frames to obtain the complete CLT.

In the microbial domain, growth behavior within colonies is often stochastic, meaning that high frame rates are required to resolve cell tracks unambiguously. While existing LAP cell trackers have been successfully used in other contexts, in practice their tracking quality deteriorates when imaging frequencies are applied that are not tuned to the colony development rate. For situations where the frame rate needs to be limited, for example to avoid exposing cells to phototoxic stress, \cite{theorell_when_2019} introduced the Uncertainty Aware Tracking (\UAT) paradigm, a multi-hypothesis cell tracking framework that incorporates knowledge of temporal cell features, such as cell elongation rates or division angles, into explainable statistical models that improve the quality of CLT inference. 
Notably, the statistical models are able to learn from past cell behavior and use this knowledge to make informed track predictions, giving the models a self-learning capacity.
Different to LAP cell trackers, \UAT{} generates a distribution of CLTs that represents the uncertainty inherent in the CLT generation process, which allows to account for this at the interpretation of the tracking results~\cite{theorell_when_2019}. However, the UAT ensemble approach comes at the cost of a higher computational effort.
 
We here present \pyUAT{}, an efficient open-source Python implementation of \UAT. 
The implementation is based on a modular design of interpretable biologically anchored statistical models that accommodate various features of cell behavior. We demonstrate the scalability of \pyUAT{} by performing cell tracking in MLCI time-lapses containing 100k+ cell instances and compare its tracking performance to those of recent LAP trackers developed for this purpose. We examine the efficacy of the statistical models and model compositions, being at the core of the UAT paradigm, in view of the tracking performance at lower imaging rates, and thereby reveal the importance to model specific cell behaviors for cell tracking.

%%%%%%%%%%%%%%%%%%%%%%%%%%%%%%%%%%%%%%%%%%%%%%
%%%%%%%%%%%%%%%%%%%%%%%%%%%%%%%%%%%%%%%%%%%%%%
\section{Approach and implementation}
\label{sec:approach}
\begin{figure*}[ht]
    \centering
    \includegraphics[width=.8\textwidth]{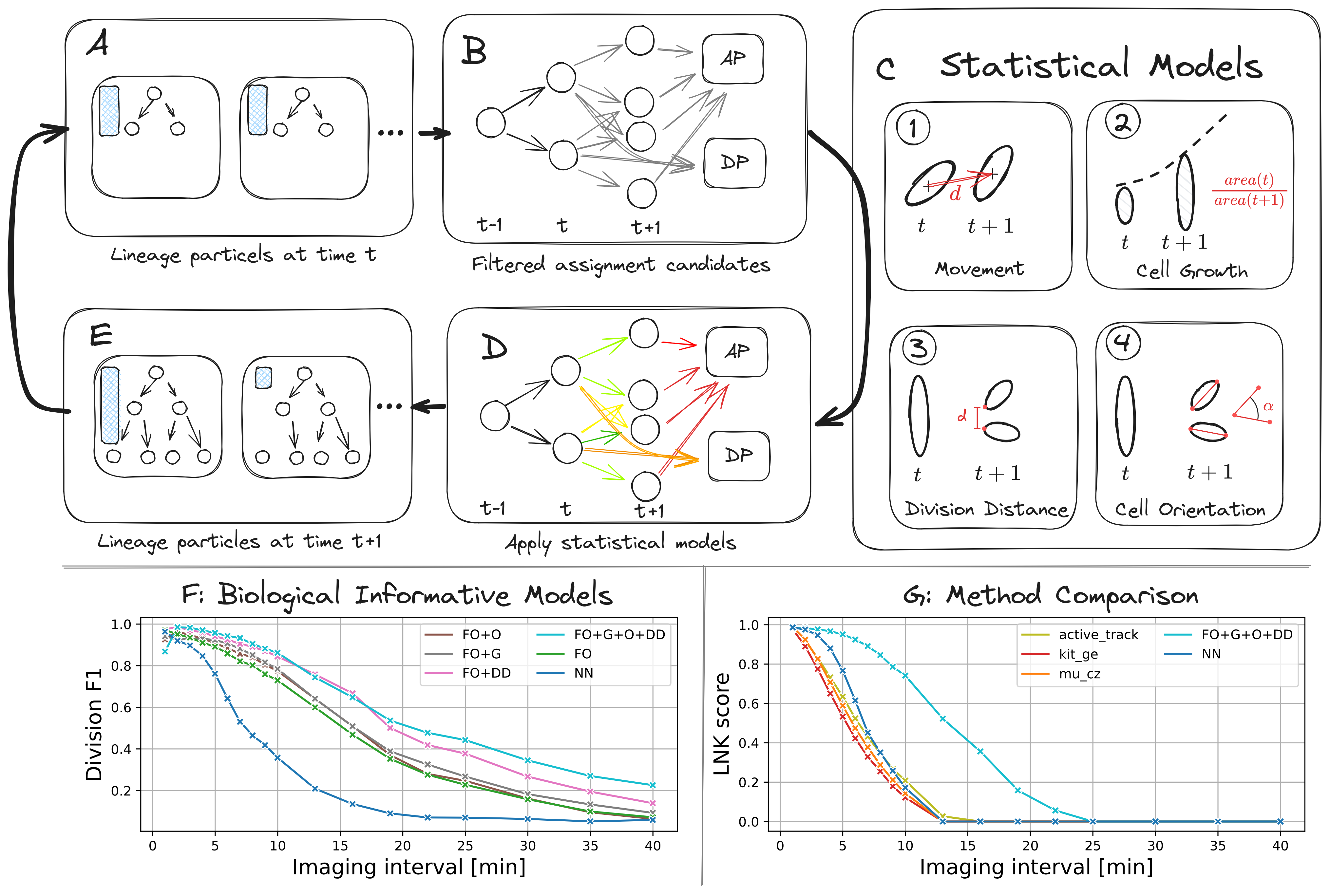}
    \caption{Schematic overview of the \pyUAT{} approach (A--E), and tracking performance comparison (F+G). 
    In (A+E), the size of the blue boxes indicates the probability of a lineage particle (the bigger, the more likely). The edge colors in B and D indicate that non-scored (grey) and probability-scored assignments (green -- high, yellow -- medium, red -- low probability). AP and DP denote appearance and disappearance assignments, respectively. 
    (F) shows \pyUAT{} derived division \textit{F1} scores for different statistical model compositions and increasing imaging intervals. 
    (G) compares the tracking quality (see text) of \pyUAT{} with established tracking methods, measuring the \textit{LNK} score at various imaging intervals (median of five time-lapse sequences).}
    % https://math.preview.excalidraw.com/#room=c1daf13ae101d5fc4876,CScuAuy-KtIIQZfLZyQ9PA
    \label{fig:workflow}
\end{figure*}

%%%%%%%%%%%%%%%%%%%%%%%%%%%%%%%%%%%%%%%%%%%%%%
%%%%%%%%%%%%%%%%%%%%%%%%%%%%%%%%%%%%%%%%%%%%%%
\UAT{} performs an iterative frame-to-frame cell tracking based on Bayesian multi-hypotheses tracking (MHT) using a particle filter (\Cref{fig:workflow}A-E). First, a distribution of CLT hypotheses (particles) is given at frame $t$ (A), where initially these particles represent empty CLTs. For each of these hypotheses, all possible assignment candidates are generated that link cells between the current ($t$) and the next frame ($t+1$) (B). Every assignment candidate is awarded a likelihood using biologically informed statistical models (C-D). Solving an integer linear program (ILP) yields the set of most likely assignments and extends the existing particles. Based on the set of most likely frame-to-frame extensions, new particles are sampled to form the updated CLT distribution at frame $t+1$, instantiating the self-learning capacity (E). This procedure is repeated for every pair of consecutive frames in the time-lapse, resembling the particle filter that finally yields the posterior probability distribution of CLTs. For the mathematical description of the Bayesian MHT approach and the particle filter, we refer to \cite{theorell_when_2019}. 

We here focus on two core elements of the \UAT{} algorithm, the formulation of the biologically informed statistical assignment models and the efficient solution of the assignment problem.
In \UAT{}, four types of assignments link cells between consecutive frames: cell appearance, disappearance, migration, and cell division. The cell appearance and disappearance assignments describe the creation of new cells and the end of cell tracks. These assignments are used to deal with cells that appear or disappear from the field of view of the image, and to deal with segmentation artefacts. The migration and cell division assignments model cell movement and division into daughter cells.
For each assignment type, we define a set of statistical models (denoted \textit{assignment model}) that score assignments according to the likelihood of known single-cell features. For example, the cell area growth model gives the likelihood of the increase in cell size within a given period of time. Further statistical models capture knowledge about cell movement, division distance, orientation. In addition a custom model can be designed in the modular \pyUAT{} framework. The four assignment models (one for each assignment type) build a \pyUAT{} tracking configuration.

Particular single-cell features are modeled using univariate probability density distributions, such as  (half-)normal distributions or kernel density estimates (Appendix~\ref{sec:assignment_models}), which are specified in \texttt{SciPy} with parameters chosen based on biological knowledge~\cite{virtanen_scipy_2020}. 
For instance, we model the single-cell area growth rate using a half-normal distribution with the colony growth rate as the mean and empirically select a variance that accounts for anticipated single-cell variation. 
A detailed guide for selecting model parameters is outlined in (Appendix~\ref{sec:modelparams}). The univariate PDF models are combined to form a joint distribution that form the assignment model. 

The particle filter then assesses assignment candidates at frame $t$ according to the likelihood provided by the underlying assignment models to sample the CLT distribution for frame $t+1$. 
Importantly and unlike existing LAP trackers, our UAT approach takes advantage of all single-cell features based on the CLT up to time $t$. This allows building powerful self-learning statistical models in a modular, Lego-like fashion that take advantage of the past cell development information, which is utilized to predict the lineage development into the future. 

Computationally, scoring assignments based on the statistical assignment models relies on the computation of single-cell features, extracted from the CLT up to frame $t$, such as the movement of a cell in past frames. Computing these features for every cell requires traversing all CLT hypotheses, and needs to be repeated for every frame-to-frame iteration. To efficiently traverse the CLT hypotheses for thousands of cells and aggregate their information along their temporal development, we developed \texttt{NumPy} array based walks through the CLTs utilizing \texttt{NumPy}'s efficient and vectorized computations (Appendix~\ref{sec:tensor_tree}). These \texttt{NumPy} arrays are efficiently distributed among parallel processes using \texttt{Ray}~\cite{Moritz2017}. To further improve efficiency, we filter for sensible assignment proposals, such as limiting the displacement radius of cells between subsequent frames. All statistical models are evaluated utilizing vectorized implementation of \texttt{SciPy} distributions.

Based on the set of scored assignments, \pyUAT{} constructs an ILP to sample likely frame-to-frame extensions. The objective function of the ILP consists of the scores of the assignment and is optimized subject to linear constraints ensuring the validity of the lineage solutions (Appendix~\ref{sec:opt_formulation}).
For solving the ILPs, proprietary (\texttt{Gurobi}, default) or open-source (\texttt{Cbc}, \url{https://github.com/coin-or/Cbc}) optimizers are available in \pyUAT{}. \texttt{Gurobi} is the default due to faster optimization performance, while \texttt{Cbc} is an open-source solution that works out of the box.
Computations are accelerated by multi-process optimization of the ILP solver or, optimally, by parallel computation of multiple CLT particles. 

%%%%%%%%%%%%%%%%%%%%%%%%%%%%%%%%%%%%%%%%%%%%%%
%%%%%%%%%%%%%%%%%%%%%%%%%%%%%%%%%%%%%%%%%%%%%%
\section{Results and discussion}
\label{sec:results}
%%%%%%%%%%%%%%%%%%%%%%%%%%%%%%%%%%%%%%%%%%%%%%
%%%%%%%%%%%%%%%%%%%%%%%%%%%%%%%%%%%%%%%%%%%%%%
We evaluate \pyUAT{} in two steps: First, we use the modular implementation to design biologically motivated models that capture typical cell behavior and investigate their importance for high-quality tracking under increasing imaging intervals. Second, we compare \pyUAT{}'s tracking quality and execution time with three recent non-DL LAP tracking methods.
For the evaluation, we use a public data set consisting of five manually curated time-lapse sequences of \Cglut{} recorded over more than 13~h, with one image taken every minute~\cite{Seiffarth2024TOIAM}. In total, the data set contains 1.4~million cell detections that are linked into more than 29k~cell tracks. To challenge the tracking methods, larger imaging intervals are generated by sub-sampling in time. 
Tracking quality is measured using the Cell Tracking Challenge (CTC, \url{https://celltrackingchallenge.net/})
\textit{LNK} score, which describes the overall quality of tracking (0 worst, 1 best), and the division \textit{F1}-score, which describes the amount of correctly reconstructed cell divisions. Both scores are computed using \texttt{traccuracy} (\url{https://github.com/Janelia-Trackathon-2023/traccuracy}).

%%%%%%%%%%%%%%%%%%%%%%%%%%%%%%%%%%%%%%%%%%%%%%
\subsection{Evaluation of tailored statistical tracking models at varying frame rates}
%%%%%%%%%%%%%%%%%%%%%%%%%%%%%%%%%%%%%%%%%%%%%%
Taking advantage of the modularity of \pyUAT{}, we build univariate models that capture specific single-cell features and assemble these models into assignment and \pyUAT{} tracking configurations to investigate their effectiveness in tracking cells at decreasing frame rates 
First, we design a baseline nearest neighbor configuration (\texttt{NN}) assuming zero cell motion and growth between consecutive images. For the second configuration, we assume that cells preserve their movement and cell area growth rate, and derive their movement and growth from cell development in the past to predict future cell positions and areas. We term this the ``first order'' model (\texttt{FO}). In both cases, we model the difference between predicted and observed cell features using a half-normal and normal distribution. Moreover, we introduce cell growth (\texttt{G}), cell orientation (\texttt{O}) and division distance assignment (\texttt{DD}) models that incorporate biological knowledge specific to the studied organism. For the \texttt{G} model, we estimate the mean single-cell area growth rate based on the colony growth (segmentation only), and model its variability using a normal distribution. The \texttt{O} model captures the ``snapping'' division behavior of \Cglut{} and models the angle between the major axes of the two daughter cells using a normal distribution. Similarly, the rotation angle between cells in a migration assignment are modeled. Finally, the division distance of daughter cells is modeled using a half-normal distribution. Details about the statistical distributions are given in Appendix~\ref{sec:assignment_models}.

\Cref{fig:workflow}F shows the \pyUAT{} tracking performance of the five configurations for a range of imaging intervals, measured using the division \textit{F1} scores. The \texttt{NN} baseline shows high division reconstruction at low imaging intervals, but the quality decreases rapidly with lower frame rates. The \texttt{FO} configuration yields much better division \textit{F1} scores, while being slightly improved by adding cell orientation (\texttt{FO+O}) or growth models (\texttt{FO+G}).
Using the division distance assignment model (\texttt{FO+DD}) enforces daughter cells of divisions to have a empirically observed close spatial distance and increases the tracking quality across a wide range of imaging intervals. Combining all models into a single composite tracking configuration (\texttt{FO+G+O+DD}) shows similar division reconstruction performance, but outperforms \texttt{FO+DD} at higher imaging intervals, thus effectively utilizing the joint information of the univariate statistical models.
The two models with the biggest improvement in division reconstruction are the \texttt{FO} and \texttt{DD}. Thus, the ability to learn information about the past cell behavior and explicitly model cell division is crucial for high-quality CLT inference.

%%%%%%%%%%%%%%%%%%%%%%%%%%%%%%%%%%%%%%%%%%%%%%
\subsection{Comparison to existing tracking methods}
%%%%%%%%%%%%%%%%%%%%%%%%%%%%%%%%%%%%%%%%%%%%%%
To investigate the tracking performance of our \pyUAT{} implementation, we select three recent non-DL tracking methods, namely MU\_CZ, KIT\_GE (Cell Tracking Challenge nomenclature) and \texttt{ActiveTrack}~\cite{loffler_graph-based_2021,ruzaeva_cell_2022}. MU\_CZ measures the overlap between segmentation masks of consecutive frames to greedily link cells (\url{https://celltrackingchallenge.net/participants/MU-CZ/}, Version: 2). KIT\_GE utilizes the graph structure of tracking and represents the tracking task as a coupled minimum cost flow problem based on cell detections~\cite{loffler_graph-based_2021}. \texttt{ActiveTrack} measures the ``activity'' of cells in consecutive images for predicting cell migration~\cite{ruzaeva_cell_2022}. All methods were used with their default parameters. 

Figure \ref{fig:workflow}G shows the tracking quality measured by the \textit{LNK} tracking score with given segmentation ground truth. The cell tracking algorithms are compared to the baseline \texttt{NN} and best-performing \texttt{FO+G+O+DD} \pyUAT{} tracking configuration. In \Cref{fig:workflow}G, we observe a strong decrease of the \textit{LNK} score for all methods at higher imaging intervals. While the \texttt{NN} configuration performs similar to three selected tracking methods and collapses at 16~min intervals, the \texttt{FO+G+O+DD} configuration consistently outperforms all other tracking methods, especially at higher imaging intervals, eventually collapsing at 25~min. Thus, the tailored statistical models and their combined biological knowledge enables \pyUAT{} to perform more robust tracking up to moderate imaging intervals.

Notably, our efficient implementation and assignment scoring makes \pyUAT{} the second fastest of the tracking methods, only slightly outperformed by the greedy MU\_CZ method (Appendix~\ref{sec:eval_tracking_models}). \pyUAT{} performs the tracking in at most $2$ hours, which is only a fraction of the recoding time of $13.3$~h.

%%%%%%%%%%%%%%%%%%%%%%%%%%%%%%%%%%%%%%%%%%%%%%
%%%%%%%%%%%%%%%%%%%%%%%%%%%%%%%%%%%%%%%%%%%%%%
\section{Conclusions}
\label{sec:conclusions}
%%%%%%%%%%%%%%%%%%%%%%%%%%%%%%%%%%%%%%%%%%%%%%
%%%%%%%%%%%%%%%%%%%%%%%%%%%%%%%%%%%%%%%%%%%%%%
The \pyUAT{} Python package is an efficient and open-source implementation of the \UAT{} paradigm. \pyUAT's modular design enables the development of advanced cell trackers by assembling tailored, interpretable statistical models that have self-learning capabilities to predict future cell behavior, a feature distinguishing the \UAT{} approach from other cell tracking methods. 
Taking advantage of the modular model composition capabilities, we have investigated the importance of each model and its impact on tracking quality and robustness, which is essential for challenging conditions such as limited frame rates.
The flexible implementation allows the adaptation and design of new assignment models beyond the studied cell organism, incorporating prior knowledge about cell behavior, making \pyUAT{} explainable and more robust compared to existing methods. The efficient implementation of \pyUAT{} performs lineage tree reconstruction in a fraction of the experiment time, making \pyUAT{} a versatile cell tracking tool.

\section*{Acknowledgments}
We acknowledge the inspiring scientific environment provided by the Helmholtz School for Data Science in  Life,  Earth  and  Energy (HDS-LEE), thank Axel Theorell for insightful discussions, and Wolfgang Wiechert for continuous support. 
This work was supported by the President's Initiative and Networking Funds of the Helmholtz Association of German Research Centres [SATOMI ZT-I-PF-04-011, EMSIG ZT-I-PF-04-44].

%%%%%%%%%%%%%%%%%%%%%%%%%%%%%%%%%%%%%%%%%%%%%%
%%%%%%%%%%%%%%%%%%%%%%%%%%%%%%%%%%%%%%%%%%%%%%
\bibliographystyle{unsrt}  
\bibliography{references}  %%% Remove comment to use the external .bib file (using bibtex).



%\bibitem{kour2014real}
%George Kour and Raid Saabne.
%\newblock Real-time segmentation of on-line handwritten arabic script.
%\newblock In {\em Frontiers in Handwriting Recognition (ICFHR), 2014 14th
%  International Conference on}, pages 417--422. IEEE, 2014.

%\bibitem{kour2014fast}
%George Kour and Raid Saabne.
%\newblock Fast classification of handwritten on-line arabic characters.
%\newblock In {\em Soft Computing and Pattern Recognition (SoCPaR), 2014 6th
%  International Conference of}, pages 312--318. IEEE, 2014.

%\end{thebibliography}
%%%%%%%%%%%%%%%%%%%%%%%%%%%%%%%%%%%%%%%%%%%%%%
%%%%%%%%%%%%%%%%%%%%%%%%%%%%%%%%%%%%%%%%%%%%%%

\newpage
\appendix
\onecolumn

{
    \centering
    \textbf{\Large 
        Supplemental Material:\\ 
        PyUAT: Open-source Python framework for efficient and scalable cell tracking with uncertainty awareness
        \vspace*{\baselineskip}\\
    }
}

%%%%%%%%%%%%%%%%%%%%%%
%%% Appendix A
\section{Scoring assignments using statistical models}
\label{sec:assignment_models}
%%%%%%%%%%%%%%%%%%%%%%
\UAT{} uses statistical models to score assignments that connect cell detections in consecutive frames using appearance, disappearance, migration and cell division assignments. Given a possible assignment of a specific type, a set of models is used to compute the assignments' likelihood. \UAT{} allows to customize the set of statistical models and tailor them to the specific behavior of the biological organism. The computed assignment likelihoods are then used in the particle filter for sampling from the lineage distribution or determining the most likely lineage.

For \UAT{} we developed various models that score appearance or disappearance of cells, their spatial movement, cell growth and orientation. In the following, we give the mathematical definitions of these models and show how to use them for computing probabilities for assignments.

%%%%%%%%%%%%%%%%%%%%%%
\subsection{Notation}
\label{sec:notation}
%%%%%%%%%%%%%%%%%%%%%%
Let $\mathbb{D}=\{D^1, \cdots, D^T\}$ be the set of all cell detections in a time-lapse of length $T \in \mathbb{N}$ where $D^t=\{d_1^t, \cdots d_{N_t}^t\}$ is the set of $N_t \in \mathbb{N}$ cell detections at frame $t$. An assignment is defined as a tuple that consists of cell detections of two consecutive frames. For the four assignment types, we denote the set of all possible assignments between frames $t$ and $t+$ as $A_{t:t+1}$. For an assignment $a \in A_{t:t+1}$, we introduce the following notation:
\begin{itemize}
    \item An appearance assignment models the appearance of a cell detection at frame $t+1$ that has no predecessor in frame $t$. We denote the assignment as
    \[a:=\left(\emptyset, \{d^{t+1}\}\right), \quad d^{t+1} \in D^{t+1}\]
    \item Similarly, the disappearance assignment contains a cell detection at frame $t$ that has no successor in frame $t+1$:
    \[a:=\left(\{d^{t}\}, \emptyset\right), \quad d^{t} \in D^t\]
    \item The migration assignment connects a cell detection at fame $t$ to a cell detection at the next frame $t+1$
    \[a:=\left(\{d^{t}\}, \{d^{t+1}\}\right), \quad d^{t} \in D^{t},\ d^{t+1} \in D^{t+1}\]
    \item The cell division assignment captures the division of a single cell into two daugther cells and connects a single cell detection at frame $t$ to two cell detections at the next frame $t+1$
    \[a:=\left(\{d^{t}\}, \{d_1^{t+1}, d_2^{t+1}\}\right), \quad d^{t} \in D^{t}, \ d_1^{t+1}, d_2^{t+1} \in D^{t+1}\]
\end{itemize}

For brevity, we omit the set-brackets from the assignment notation. For instance, a cell division assignment is expressed as
\begin{equation}
    a := \left( d^t; d_1^{t+1}, d_2^{t+1} \right)
\end{equation}

Furthermore, $\phi_{norm}$ and $\phi_{hn}$ denote the probability density function (PDF) of a normal and half-normal distribution, respectively.

\subsection{Constant probability model (CP)}
\label{sec:model_cp}

The simplest model for scoring assignments is a constant probability model. 
Without extracting a single-cell quantity, this model will always return a constant probability.
\begin{equation}
    p(a) = c, \quad c \in [0,1]
\end{equation}
where $a$ is an assignment and $c$ is the constant probability.
We use the model mainly to score appearance and disappearance assignments. These events happen when cells leave the microscopy's field of view or due to errors in the segmentation.

%%%%%%%%%%%%%%%%%%%%%%
\subsection{Nearest neighbor model (\texttt{NN})}
\label{sec:stay_models}
%%%%%%%%%%%%%%%%%%%%%%
For slow cell growth behavior, where cells do not move much nor change their size between consecutive images, an assignment model
should give high probability to assignments where cells stay at the same position and  maintain their size (nearest neighbor). For migration assignments, we directly compare the position and cell size, whereas for division assignments, we compute the center of mass and joint size to compare it with the position and cell size of the previous frame. We utilize a half-normal distribution with mode 0 (no movement) to model the distance between linked cells and a positive scale parameter defining the variance of the distribution. The relative size change for the growth model is scored using a normal distribution with a mean of 1 (no growth) and a positive scale parameter.

Thus, for the movement and growth models we compute the likelihood for an assignment $a=(d^t, d^{t+1}) \in A_t$:
\begin{equation}
    p(d^t, d^{t+1}) = \phi_{hn}\left(dist\left(d^t, d^{t+1}\right)\right) \cdot \phi_{norm}\left(\frac{area(d^{t+1})}{area(d^t)}\right)
\end{equation}
where $dist(\cdot, \cdot)$ denotes the Euclidean distance of the two linked detections and $area(\cdot)$ the size of the detection, respectively. For positions, this is the Euclidean distance between the centers of the two cell detections, and for cell sizes, this is the relative area change (growth).

Similarly, for cell division models, we define
\begin{equation}
    p(d^t; d_1^{t+1}, d_2^{t+1}) = 
    \phi_{hn}\left(dist\left(d^t; d_1^{t+1}, d_2^{t+1}\right)\right) \cdot \phi_{norm}\left(\frac{area(d_1^{t+1}) + area(d_2^{t+1})}{area(d^t)}\right)
\end{equation}
where $dist(\cdot; \cdot, \cdot)$ denotes the Euclidean distance of the parent cell to the center of mass of the progeny.

%%%%%%%%%%%%%%%%%%%%%%
\subsection{First-order models (\texttt{FO}): Predict movement \& Predict growth}
\label{sec:model_fo}
%%%%%%%%%%%%%%%%%%%%%%
\UAT{} has access to the complete CLT and the temporal evolution of cell properties during its iterative tracking procedure. The prediction models utilize this knowledge about the temporal development of single-cell quantities up to frame $t$ for predicting their development for the frame $t+1$.

The residual between a predicted single-cell quantity and a actually observed cell quantity at time $t+1$ is computed and scored using a statistical distribution. Both quantities (position and area) are scored using a half-norm distribution with mode 0 (exactly predicted cell position/size) and a positive scale parameter.

Thus, similar to the \texttt{NN} model, we write the migration probability
\begin{equation}
    p(d^t, d^{t+1}) = \phi_{hn}\left(dist(pred(d^t), d^{t+1})\right)
\end{equation}
and cell division probability
\begin{equation}
    p(d^t; d_1^{t+1}, d_2^{t+1}) = \phi_{norm}\left(dist\left(pred(d^t); d_1^{t+1}, d_2^{t+1}\right)\right)
\end{equation}
where $pred(\cdot)$ predicts the cell property at frame $t+1$, and $dist(\cdot, \cdot)$ or $dist(\cdot; \cdot,\cdot)$ computes the element-wise distance in the cell's feature space. The prediction of cell property is performed by averaging the cell property over a time span of $n \in \mathbb{N}$ frames. Therefore, we perform walks of length $n \in \mathbb{N}$ along the cell lineage tree (CLT) using the \texttt{tensortree} library (see Appendix~\ref{sec:eval_tracking_models}). Longer walks may provide more robust predictions while short walks prioritize the most recent development. In this work, we used short walks of $n=1$. Notice, that for a walk length of $n=0$, the models are equivalent to the no movement/growth models in Sec. Nearest neighbor model.

%%%%%%%%%%%%%%%%%%%%%%
\subsection{Orientation models (O)}
\label{sec:model_orientation}
%%%%%%%%%%%%%%%%%%%%%%
Living \Cglut{} cells usually change their orientation only slowly throughout their development. We quantify this change in orientation by the angle between the major-axis of the rod-shaped cells in consecutive images. For cell migration the angle changes slowly due to cell growth and displacement and is modeled using a half-normal distribution with a mean of 0 and a positive scale parameter. However, during cell division, \Cglut{} exhibits a ``snapping'' behavior such that the daughter cells end up in a characteristic angle towards each other. Therefore, for cell division events, we model the angle of the daughter cells using a normal distribution with a mean of 135 degrees and a positive scale parameter.

Thus, the likelihood of a migration assignment is defined as
\begin{equation}
    p(d^t, d^{t+1}) = \phi_{norm}\left (angle\left (d^{t}, d_{t+1}\right) \right)
\end{equation}
and for a cell division assignment by
\begin{equation}
    p(d^t; d_1^{t+1}, d_2^{t+1}) = \phi_{norm}\left( angle \left (d_1^{t+1}, d_2^{t+1} \right) \right)
\end{equation}
where $\text{angle}(\cdot, \cdot)$ denotes the angle between the major axes of two cell detections.

%%%%%%%%%%%%%%%%%%%%%%
\subsection{Division distance model (\texttt{DD})}
\label{sec:model_mitosis_distance}
%%%%%%%%%%%%%%%%%%%%%%
When cells divide into two daughter cells, the progeny are usually very close together, with no cell between them. We utilize this observation and model the distance between daughter cells using a half-norm distribution with mode 0 and a positive scale parameter.
\begin{equation}
    p(d^t; d_1^{t+1}, d_2^{t+1}) = \phi_{norm}\left( dist \left( d_1^{t+1}, d_2^{t+1} \right) \right)
\end{equation}
where $dist(\cdot, \cdot)$ denotes the minimum Euclidean distance between the major axes of two cell detections.

%%%%%%%%%%%%%%%%%%%%%%
\subsection{Biological growth model (\texttt{G})}
\label{sec:model_bio_growth}
%%%%%%%%%%%%%%%%%%%%%%
For microbial organisms, usually at least some information about their growth capabilities is available for the studied cultivation conditions. However, if this is not the case, we can still utilize the segmentation information to quantify the growth rates of the microbial colony and use this information as an estimate for the single-cell growth rates. In this work, we choose mean values of $1.008$ and $1.016$ as the expected growth rate for migration and cell division assignments, respectively. Using a normal distribution with these mean values, the probability of the assignments using the growth model is computed as follows for migration
\begin{equation}
    p(d^t, d^{t+1}) = \phi_{norm}\left(\frac{area(d^{t+1})}{area(d^t)}\right)    
\end{equation}
and division assignments
\begin{equation}
    p(d^t; d_1^{t+1}, d_2^{t+1}) = \phi_{norm}\left(\frac{area(d_1^{t+1}) + area(d_2^{t+1})}{area(d^t)}\right)
\end{equation}
When subsampling the images, the temporal distance between the images is increased and larger frame-to-frame cell size growth is to be expected between consecutive frames. In our exponential growth model, we compensate by introducing an exponential subsampling factor. For a subsampling factor $\tau \in \mathbb{N}$, we choose mean $\bar \mu_{\text{migration}}$ and $\bar \mu_{\text{division}}$ for migration and cell division assignments
\begin{equation}
    \bar \mu_{\text{migration}} = 1.008^\tau, \quad
    \bar \mu_{\text{division}} = 1.016^\tau
\end{equation}

%\newpage
%%%%%%%%%%%%%%%%%%%%%%
\subsection{Computing probabilities from densities}
\label{sec:compute_prob}
%%%%%%%%%%%%%%%%%%%%%%
The half-normal or normal distributions used for assignment models do not provide probabilities, but probability densities. For computing probabilities, we need to integrate the probability density function (PDF). Therefore, we define the probability of a measured quantity (e.g. size change, position difference) as the probability of more extreme quantities. Let $q\in \mathbb{R}$ be the quantity extracted for a specific assignment $a$, and let $\phi(\cdot)$ be the PDF for an assignment model. Then the probability of assignment $a$ is defined as:
\begin{equation}
    p(q) = 1 - \int_{M-\delta}^{M+\delta} \phi(x; \theta)~dx
\end{equation}
where $M$ is the mode of the distribution, $\theta$ are distribution parameters that define the distribution's PDF, and $\delta = |M-q|$ is the distance between mode and the measured quantity.

%%%%%%%%%%%%%%%%%%%%%%
\subsubsection{Half-normal densities}
%%%%%%%%%%%%%%%%%%%%%%
For half-normal PDFs $\phi_{hn}(\cdot)$ with mode $M=0$, scale parameter $\sigma \in \mathbb{R}_{> 0}$, and observed quantity $q \in \mathbb{R}_{\geq 0}$, the probability is computed as:
\begin{eqnarray}
    \phi_{hn}(q) & =& 1 - \int_{M-\delta}^{M+\delta} \phi_{hn}(x; \sigma)~dx = 1 - \int_{0-|0-q|}^{0+|0+q|} \phi_{hn}(x; \sigma)~dx = 1 - \int_{-q}^{+q} \phi_{hn}(x; \sigma)~dx \nonumber \\
    & =& 1 - \int_0^q \phi_{hn}(x; \sigma)~dx.
\end{eqnarray}

%%%%%%%%%%%%%%%%%%%%%%
\subsubsection{Normal densities}
%%%%%%%%%%%%%%%%%%%%%%
For normal PDFs $\phi_{norm}(\cdot)$, mean $M=\mu$, scale parameter $\sigma \in \mathbb{R}_{> 0}$, and observed quantity $q \in \mathbb{R}$, the probability is computed as:
\begin{eqnarray}
    p_{norm}(q) & =& 1 - \int_{M-\delta}^{M+\delta} \phi_{norm}(x; \sigma)~dx = 1 - \int_{\mu - \delta}^{\mu + \delta} \phi_{norm}(x; \sigma)~dx \nonumber \\
    & =& \int_{-\infty}^{\mu - \delta} \phi_{norm}(x; \sigma)~dx + \int_{\mu + \delta}^{\infty} \phi_{norm}(x; \sigma)~dx.
\end{eqnarray}

\newpage
%%%%%%%%%%%%%%%%%%%%%%
%%% Appendix B
\section{Choosing model parameters - Tuning knobs for single-cell behavior modeling}
\label{sec:modelparams}
\begin{figure}[h]
    \centering
    %\includesvg[width=0.9\linewidth]{scale_search.svg}
    \includegraphics[width=1.0\linewidth]{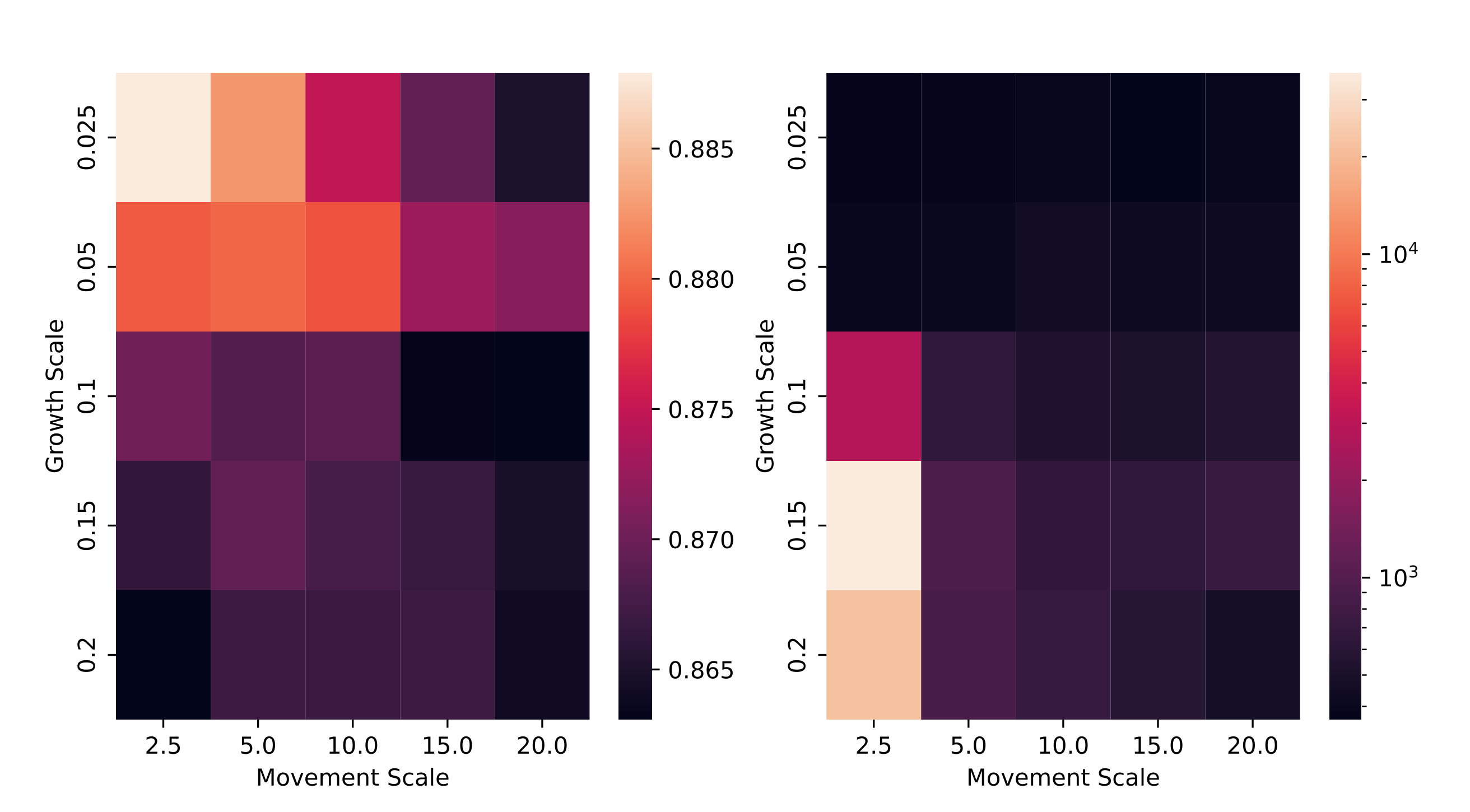}
    \caption{Grid search results for growth and movement scale parameters showing TRA score results (left) and tracking execution time in seconds (right).}
    \label{fig:scale_grid_search}
\end{figure}

%%%%%%%%%%%%%%%%%%%%%%
\UAT{} utilizes tailored models to rate assignments, so that reasonable ones are preferred over nonsensical ones, and combines them in frame-to-frame tracking by utilizing a particle filter approach to iteratively fuse them into full CLTs. The tailored assignment models presented in Appendix~\ref{sec:assignment_models} allow for statistical modeling of single-cell properties of microbes, for example, their movement or growth behavior. However, all these statistical models and their PDFs come with parameters such as scale or mode. While the mode parameter can be derived from intuition (e.g., no movement) or biological prior knowledge (e.g., expected growth), the scale parameter indicates the spread of the distribution. To investigate the sensitivity of the tracking result derived by \UAT{} with respect to the choice of the these scale parameters, we selected the first-order tracking configuration and performed a grid search over the scale parameter for the movement and growth models. \Cref{fig:scale_grid_search} shows the median \texttt{TRA} score and execution time evaluated with an imaging interval of $10$ minutes. The heatmap shows that the tracking quality is very similar for a wide range of scale parameters. However, the smallest growth and movement scale parameters provide the best tracking result in terms of the \texttt{TRA} score. This indicates that the models we designed fit well to the single-cell properties of the studied organism.

Also, the execution time remains very similar over a wide range of scale parameters, except of the region of very low movement scale and high growth scale parameters. 

Based on these results, we recommend users to start with large scale parameters and refine them according to the tracking results. This iterative procedure can help to find appropriate parameters tailored to other microbes or cultivation conditions. Manual annotation corrections help to build ground truth data sets to evaluate the tracking performance and find suitable parameters.

\newpage
%%%%%%%%%%%%%%%%%%%%%%
%%% Appendix C
%%%%%%%%%%%%%%%%%%%%%%
\section{Tensor Walks: Vectorized walks in cell lineage trees}
\label{sec:tensor_tree}
\begin{figure}[h]
    \centering
    \includegraphics[width=\linewidth]{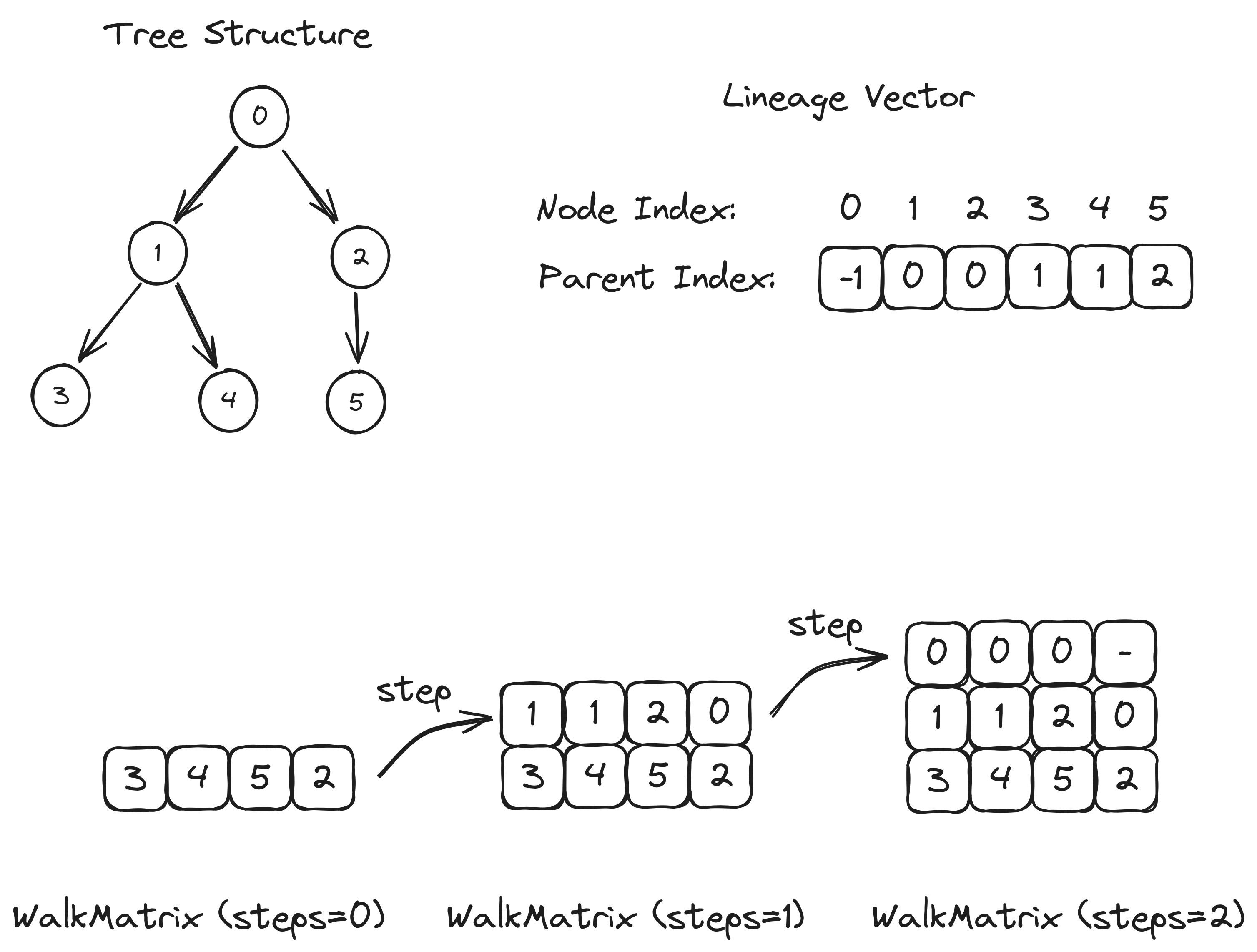}
    \caption{Lineage tree example (top left) and its vectorized CLT representation (top right). The nodes are labeled with indices. The lineage vector stores the inverse links of the CLT (left), i.e., for every node the parent index is stored (or -1 if no parent is available). The bottom shows the exemplary walk matrix generation starting at nodes $3,4,5,2$ and performing for two steps. Walks can be reconstructed by traversing the columns of the walk matrix.}
    \label{fig:vectorized_tree}
\end{figure}

%%%%%%%%%%%%%%%%%%%%%%
During the \UAT{} tracking procedure, quantities need to be extracted from CLTs to score assignment candidates by the assignment models. Therefore, during the execution \UAT{} often needs to compute walks on the CLT and derive node associated single-cell property development such as position or cell size. To perform the walks in the tree structure efficiently, we developed the \twalks{} library that executes vectorized tree walks on a vectorized tree format (see \Cref{fig:vectorized_tree}).

To transfer the CLT into its vectorized form, we first label all nodes with indices, build a tree vector and store the parent of the nodes (or -1 if no parent available). This format allows performing walks within the tree using simple index look-ups in \texttt{numpy} and, therefore, makes use of vectorized CPU instructions. Walks are performed in a so called \textit{walk matrix} (see \Cref{fig:vectorized_tree}). We initialize the first row of the walk matrix with the start node indices. With every step, a walk towards the predecessor node is performed using index look-ups in the vectorized CLT. This is continued for a fixed number or until all walks stopped (e.g., due to specified criteria).

We compare the \twalks{} execution performance to the widely used and general \texttt{networkx} library (\url{https://github.com/networkx/networkx}). We evaluate the task of estimating the age of a node, that is, the length of its walk to its first predecessor that has a sibling. \twalks{} utilizes the vectorized walk computation and \UAT{} uses standard python loops for computing the age of nodes in a \texttt{networkx} graph structure. First benchmarks in sequential trees (i.e., every node index links to the previous index) show a four to five  fold acceleration in execution time using the \twalks{} library compared to \texttt{networkx}, for computing the age of a node in the tree.

\newpage
%%%%%%%%%%%%%%%%%%%%%%
%%% Appendix D
\section{Computing optimal frame-to-frame lineages using integer linear programming}
\label{sec:opt_formulation}
%%%%%%%%%%%%%%%%%%%%%%

\UAT{} utilizes a particle filter approach to iteratively combine frame-to-frame lineages into a CLT that covers the full time-lapse. To compute the frame-to-frame lineages \UAT{} uses an integer linear program (ILP) formulation. In the following we describe the construction of the frame-to-frame ILP.

Let $A_{t:t+1} = \{a_1, a_2, \ldots\}$ be the set of assignment candidates that are considered for connecting cell detections $D_t=\{d^t_1, \ldots, d^t_{N_t}\}$ at frame $t$ to cell detections $D_{t+1}=\{d_1^{t+1}, \ldots, d_{N_{t+1}}^{t+1}\}$ at the next frame $t+1$. The likelihood of an assignment $a \in A_{t:t+1}$ is computed by
\begin{equation}
    p(a) = \prod_{m \in MS(a)} p_m(a) \quad \in [0,1]
\end{equation}
where $MS(a)=\{m_1(\cdot), m_2(\cdot), \ldots\}$ is the set of assignment models with probability functions $p_m(\cdot)$ that are used for rating the type of assignment (see Appendix~\ref{sec:assignment_models}).

To represent the selection of all assignments, we define the boolean vector $\Vec{b} \in \{0,1\}^{|A_{t:t+1}|}$ that contains a boolean entry for every assignment candidate between frames $t$ and $t+1$, where an entry of $1$ denotes that the assignment is part of the selection $\Vec{b}$, whereas a value of $0$ indicates that the assignment is not part of the selection $\Vec{b}$. The selection $\Vec{b}$ represents a valid frame-to-frame tracking, if and only if every detection at frame $t$ and $t+1$ is part of exactly one selected assignment. This is expressed in terms of linear constraints
\begin{eqnarray}
    \label{eq:valid_lineage_1}
    \forall d^t \in D_t &:& \sum_{i=0}^{|A^{t:t+1}|} \mathbbm{1}(d^t, a_i) = 1 \\
    \label{eq:valid_lineage_2}
    \forall d^{t+1} \in D_{t+1}&:& \sum_{i=0}^{|A^{t:t+1}|} \mathbbm{1}(d^{t+1}, a_i) = 1
\end{eqnarray}
where $\mathbbm{1}(\cdot, \cdot) \in \{0,1\}$ indicates that detection $d$ is part of assignment $a$ or not:
\begin{equation}
    \mathbbm{1}(d,a)=\begin{cases}
      1, & \text{if detection $d$ is part of assignment $a$} \\
      0, & \text{otherwise}
    \end{cases}
\end{equation}

The most likely frame-to-frame tracking $\Vec{b}_{opt}$ is computed by maximizing the joint probability of the selected assignments, while satisfying the conditions in Eqs.~\eqref{eq:valid_lineage_1}-\eqref{eq:valid_lineage_2}:
\begin{equation}
    \Vec{b}_{opt} = \operatorname*{argmax}_{\Vec{b} \in \{0,1\}^{\left|A_{t:t+1}\right|}} \prod_{i=0}^{|A_{t:t+1}|} p(a_i)^{b_i}.
\end{equation}

By applying the logarithm to the objective function, we reformulate the optimization problem in terms of a linear objective function
\begin{equation}
     \Vec{b}_{opt} = \operatorname*{argmax}_{\Vec{b} \in \{0,1\}^{\left|A_{t:t+1}\right|}} \prod_{i=0}^{|A_{t:t+1}|} p(a_i)^{b_i} = \operatorname*{argmax}_{\Vec{b} \in \{0,1\}^{\left|A_{t:t+1}\right|}} \sum_{i=0}^{|A_{t:t+1}|} b_i \cdot \log p(a_i).
\end{equation}
We solve this ILP for the optimal frame-to-frame tracking $\Vec{b}_{opt}$ using \texttt{Gurobi} or \texttt{CBC}.

\newpage
%%%%%%%%%%%%%%%%%%%%%%
%%% Appendix E
\section{Evaluating tracking models}
\label{sec:eval_tracking_models}
%%%%%%%%%%%%%%%%%%%%%%
Evaluating the performance of different tracking  configurations requires the comparison of ground truth lineages (GTL) and predicted cell lineages (PCL). In the latter, we introduce different scoring schemes to asses the alignment of the PCL and GTL, measure the scores and execution time of the algorithms at different imaging intervals, and evaluate the efficiency of the introduced models for \UAT{}.

\subsection{Tracking configurations}

The assignment models described in Appendix~\ref{sec:assignment_models} compute likelihoods for a specific property of the assignment, for instance, the cell movement or growth rate. To score multiple properties, we define sets of these assignment models for every type of assignments, that is, appearance and disappearance assignments, migration and cell division assignments. \Cref{tab:model_sets} gives an overview of the designed tracking configurations and the models involved for computing assignment likelihoods. The \texttt{NN} configuration utilizes the constant probability model to score appearance and disappearance assignments (used in all other configurations, too). The migration and cell division assignments are scored with the ``no movement'' and ``no growth'' models. The first-order (\texttt{FO}) configuration utilizes the first-order models to predict and score movement and cell size changes (see Appendix~\ref{sec:assignment_models}). We utilize the \texttt{FO} model as a baseline and combine all the following configurations with this configuration. The \texttt{FO}+orientation model adds the orientation models to the configuration. The \texttt{FO}+distance model adds the cell division distance model to the configuration, and \texttt{FO}+growth model overwrites the \texttt{FO} growth model and utilizes the growth model based on biological prior knowledge.

\begin{table}[h]
    \centering
    \footnotesize
    \caption{Designed \UAT{} tracking configurations. All configurations define four sets of models that compute likelihoods for specific type of assignment. Values in brackets indicate the mode/mean and scale parameters of the statistical distributions (single value indicates zero mode/mean and only specifies the scale parameter). Tracking configurations that are combined with ``\texttt{FO} only'' show the additional models. Appearance and disappearance models are not shown, but get a value of $0.25$ per assignment. The subsampling rate $\tau$ is used to adapt the statistical distribution parameters.}
    \begin{tabular}{|c|c|c|}
        \hline
        Tracking Configuration & migration assignment & cell division assignment \\
        \hline
        Nearest Neighbor (\texttt{NN}) & \makecell{No movement: half-norm($20 \cdot \tau$)\\No growth: norm($0.05 \cdot \tau$)} & \makecell{No movement: half-norm($20 \cdot \tau$)\\No growth: norm($0.1 \cdot \tau$)} \\
        \hline
        First-Order (\texttt{FO}) & \makecell{Predict movement: half-norm($20 + 5 \cdot \tau$)\\Predict size: norm($50 + 10 \cdot \tau$)} & \makecell{Predict movement: half-norm($20 + 2 \cdot 5 \cdot \tau$)\\Predict size: norm($50 + 2 \cdot 10 \cdot \tau$)} \\
        \hline
        \texttt{FO}+orientation (\texttt{FO+O}) & \makecell{Migration angle: half-norm($20 \cdot \tau$)} & \makecell{Division angle: norm(135, $20 \cdot \tau$)} \\
        \hline
        \texttt{FO}+distance (\texttt{FO+DD}) & - & \makecell{Division distance: half-norm(3)} \\
        \hline
        \texttt{FO}+growth (\texttt{FO+G}) & \makecell{Growth model: norm($1.008^s$, $0.05 \cdot \tau$)} & Growth model: norm($1.016^s$, $0.1 \cdot \tau$) \\
        \hline
    \end{tabular}
    \label{tab:model_sets}
\end{table}

\newpage
\subsection{Execution time}

Different tracking configurations have different runtimes for deriving the CLT. 
The execution times of the tracking configurations were measured on a system equipped with 2x AMD EPYC 7282 16-core processor and 504 GB. All tracking methods were executed on a single core, while the evaluation of the different tracking methods and imaging intervals was run in parallel batches of 32. The execution time includes data loading, storing and the tracking runtime.

\begin{figure}
    \centering
    \includegraphics[width=1.\linewidth]{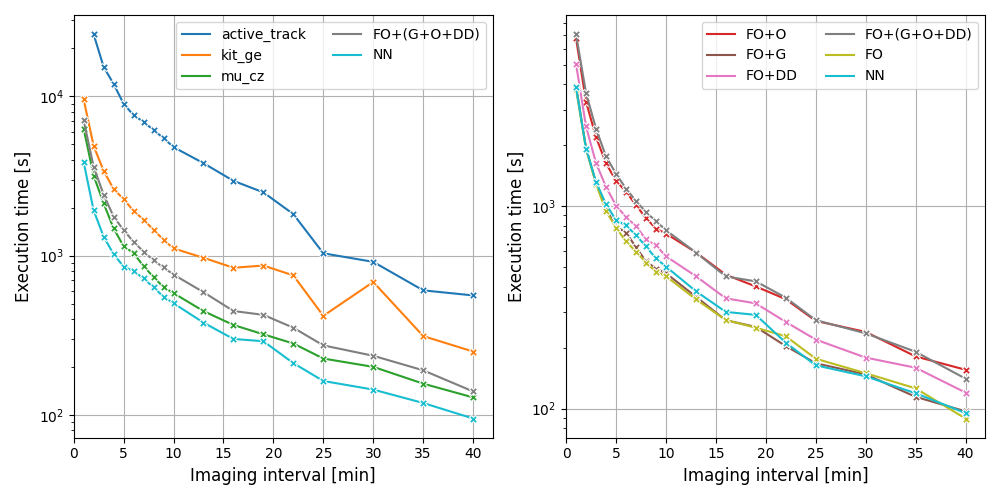}
    \caption{Execution time of \UAT{} in comparison with the existing tracking methods (left) and among the different tracking configurations (right) at various imaging intervals.}
    \label{fig:exec_time}
\end{figure}

\Cref{fig:exec_time} shows the measured execution times for all five \UAT{} tracking configurations in \Cref{tab:model_sets} and were compared with those of existing tracking methods, \texttt{MU\_CZ}, \texttt{KIT\_GE} and \texttt{ActiveTrack}. We find that \UAT{} is among the fastest of the tracking methods and that more complex statistical models only lead to a slight increase in the execution time. Moreover, lower frame rates generally lead to a reduction in execution time, as the tracking needs to be performed for fewer images and, therefore, fewer cell detections.

\end{document}